\documentclass[preprint,prd,noshowpacs,nofootinbib]{revtex4}

\usepackage{color} 
\usepackage{amssymb}
\usepackage{amsmath}
\usepackage{graphicx} 
\usepackage{float}
\usepackage{braket}     
\usepackage{pdfpages}
\usepackage{appendix}
\usepackage{epstopdf}
\usepackage[utf8]{inputenc}

\begin{document}

\title{GUP, Lorentz Invariance (Non)-Violation, and\linebreak Non-Commutative Geometry}

\author{Michael Bishop}
\email{mibishop@mail.fresnostate.edu}
\affiliation{Mathematics Department, California State University Fresno, Fresno, CA 93740, USA}

\author{Daniel Hooker}
\email{dshooker@mail.fresnostate.edu}
\affiliation{Department of Physics, California State University Fresno, Fresno, CA 93740, USA}

\author{Peter Martin}
\email{peter.martin.1@vanderbilt.edu}
\affiliation{Vanderbilt University, PMB 401807, 2301 Vanderbilt Place, Nashville, TN 37240, USA}

\author{Douglas Singleton}
\email{dougs@mail.fresnostate.edu}
\affiliation{Department of Physics, California State University Fresno, Fresno, CA 93740, USA}

\date{\today}

\begin{abstract}
In this work, we formulate a generalized uncertainty principle with both position and momentum operators modified from their canonical forms. We study whether Lorentz symmetry is violated and whether it can be saved with these modifications. The requirement that Lorentz invariance is not violated places restrictions on the way the position and momentum operators can be modified. We also investigate the connection between general uncertainty principle and non-commutative geometry models, e.g.,laying out the connection between area/area operators and angular momentum in both models.
\end{abstract}

\maketitle
 


\section{Introduction}

The generalized uncertainty principle (GUP) and its associated minimal length/maximum energy--momentum scale provide a potentially testable approach to quantum gravity \cite{vene, gross,gross2,amati2,amati3,amati4,amati,maggiore,garay,KMM,scardigli,adler-1999}. The GUP approach to quantum gravity has led to various proposed table-top laboratory experiments of quantum gravity effects \cite{vagenas,vagenas2,bekenstein,bekenstein-2}.

GUP models can lead to the violation of Lorentz invariance, which in turn can give rise to observable consequences such as photon dispersion from gamma-ray bursts \cite{AC-nature}. The lack of observation of photon dispersion from several powerful gamma-ray bursts in 2009~\cite{grb-1} and in 2024 \cite{grb-2} pushed the quantum gravity scale past the Planck scale. This is with the assumption that in quantum gravity, photon dispersion comes in at a linear order. For a quadratic effect, the bounds are not as stringent.
One of the aims of this paper is to construct a GUP model with minimal length, which nevertheless is Lorentz-invariant. The violation of Lorentz symmetry can be connected to some models of dark matter \cite{dm} or dark energy~\cite{de}.

In addition to preserving the Lorentz invariance structure in the GUP models proposed here, we also want to maintain the standard commutation relationship between the modified position and modified momentum operators, i.e., $[{\hat x}, {\hat p}] = i \hbar$. We find that while it is possible to preserve both Lorentz invariance and the standard commutation relationship in one spatial dimension (1D), it is no longer possible to preserve both once one moves to three spatial dimensions (3D). 

Another aim of this work is to connect the GUP approach to quantum gravity with the non-commutative geometry approach to quantum gravity \cite{carroll,smailagic,spallucci}. GUPs in three spatial dimensions can exhibit non-commutative coordinates, whereas GUPs in one spatial dimension do not. Since many GUP models focus on only one spatial dimension, this connection between GUPs and non-commutative geometry has not been deeply studied. There are exceptions, including \cite{KMM}, which considered a GUP in three spatial dimensions, or more recent work \cite{todrinov}, which studied a $3+1$-dimensional relativistic GUP.  

\section{GUP Models in One Spatial Dimension}\label{sec2}

In this section, we construct a GUP in one spatial dimension along the lines sketched out in the introduction, i.e., we build modified position and momentum operators which preserve the standard quantum mechanical commutation relationships. In the rest of this work, capital letters indicate modified operators, and small letters are the standard/canonical operators.

We begin by writing a general commutation relationship in one spatial dimension as 
\begin{equation}
[{\hat X} , {\hat P}] = i \hbar F\left(\frac{{\hat x}}{x_m}, \frac{{\hat p}}{p_M} \right)~.
\label{eq:modcomm}
\end{equation}
The capital ${\hat X}$ and ${\hat P}$ are the modified 
position and momentum, and  the general function  $F\left(\frac{{\hat x}}{x_m}, \frac{{\hat p}}{p_M} \right)$ depends on the standard position and momentum, ${\hat x}$ and ${\hat p}$, which satisfy the standard commutator $[{\hat x} , {\hat p}] = i \hbar $. 
The constants $x_m$/$p_M$ set the small distance/large momentum cut-off scale. 
When $\Delta x \gg x_m$ and/or $\Delta p \ll p_M$, the ansatz function should have the limit $ F\left(\frac{{\hat x}}{x_m}, \frac{{\hat p}}{p_M} \right) \to 1$, i.e., the modified commutator becomes the standard one for large distances and/or small momenta \cite{plb-2023}. 

Next, we need to choose ${\hat X}$ and ${\hat P}$ to obtain \eqref{eq:modcomm}. A general form for the modified position and momentum operator are
\begin{equation}
    \label{xp1d}
{\hat X} = i \hbar G\left(\frac{{\hat x}}{x_m}, \frac{{\hat p}}{p_M} \right) \partial _{p}  ~~~~ {\rm and} ~~~~ {\hat P} =  H\left(\frac{{\hat x}}{x_m}, \frac{{\hat p}}{p_M} \right) ~{\hat p}~,
\end{equation}
where the ansatz functions $G$ and $H$ are chosen to give  \eqref{eq:modcomm}. We want to select $G$ and $H$ so that $F=1$ while at the same time still giving a minimum distance. We simplify the analysis by taking all the ansatz functions to depend only on the momentum, i.e., $G\left(\frac{{\hat x}}{x_m}, \frac{{\hat p}}{p_M} \right) \to G\left(\frac{{ p}}{p_M} \right)$ and $H\left(\frac{{\hat x}}{x_m}, \frac{{\hat  p}}{p_M} \right) \to H\left(\frac{{ p}}{p_M} \right)$, where we have let ${\hat p} \to p$ since in momentum space the momentum operator is simply multiplication by $p$.

For the ansatz function $H$, we want ${\hat P} \approx p$ for $p \ll p_M$ and ${\hat P} \approx p_M$ for $p \gg p_M$. In references \cite{BJLS,grf-2022}, two specific forms of ansatz function $H$ which satisfy these conditions were given as
\begin{equation}
    \label{ansatz-H}
  H\left( \frac{p}{p_M} \right) = \frac{2 p_M}{\pi p} \arctan \left( \frac{\pi p}{2 p_M} \right) ~~~~{\rm and}~~~~
    H\left( \frac{p}{p_M} \right) = \frac{p_M}{p} \tanh \left( \frac{p}{p_M} \right) ~.
\end{equation}
Once the ansatz function $H$ is defined, the ansatz function $G$ is determined by requiring that the ansatz function in the commutator \eqref{eq:modcomm} is $F=1$, i.e.,
\begin{equation}
    \label{cond1}
    G\left( \frac{p}{p_M} \right) \partial_p \left[ H\left(\frac{p}{p_M} \right)~p\right] =1. 
\end{equation}
Using \eqref{cond1} and the two ansatz functions in \eqref{ansatz-H}, the functions for $G$ are given by 
\begin{equation}
    \label{ansatz-G}
    G\left( \frac{p}{p_M} \right) = \left[ 1+ \left( \frac{\pi p}{2 p_M}\right)^2 \right] 
    ~~~~{\rm and}~~~~
    G\left( \frac{p}{p_M} \right) = \cosh ^2 \left( \frac{p}{p_M}\right)   ~,
\end{equation}
where the first $G$ is for $H \sim \arctan$ and the second $G$ is for $H \sim \tanh$.

Using the two sets of ansatz functions from \eqref{ansatz-H} and \eqref{ansatz-G} in \eqref{xp1d}, we find that the modified position and momentum operators are
\begin{equation}
    \label{xp-arctan}
{\hat X} = i \hbar \left[ 1+ \left( \frac{\pi p}{2 p_M}\right)^2 \right] \partial _{p}  ~~~~ {\rm and} ~~~~ {\hat P} =  \frac{2 p_M}{\pi} \arctan \left( \frac{\pi p}{2 p_M} \right) ~,
\end{equation}
and 
\begin{equation}
    \label{xp-tanh}
{\hat X} = i \hbar \cosh ^2 \left( \frac{p}{p_M}\right) \partial _{p}  ~~~~ {\rm and} ~~~~ {\hat P} =  p_M \tanh \left( \frac{p}{p_M} \right) ~,
\end{equation}
The modified operators in \eqref{xp-arctan} and \eqref{xp-tanh} both lead to the standard commutator  $[{\hat X}, {\hat P}] = i \hbar$ \cite{BJLS,grf-2022}. Therefore, the standard uncertainty also remains the same $\Delta X \Delta P \ge \frac{\hbar}{2}$, but now in terms of the modified position and momentum. Since the momentum operator is capped at $p_M$  for both \eqref{xp-arctan} and \eqref{xp-tanh}, we find that $\Delta P \sim p_M$ for both cases in \eqref{xp-arctan} and \eqref{xp-tanh}. Thus, the minimum distance for both versions of the modified operators is $\Delta X \Delta P \ge \frac{\hbar}{2} \to \Delta X \ge \frac{\hbar}{2 p_M}$.

\section{Generalizing to $3+1$ Dimensions}\label{sec3}

We now generalize the results of the previous section to $3+1$ space--time dimensions. To make this transition, the modified commutator from \eqref{eq:modcomm} now becomes  
\begin{equation}
[{\hat X}_\mu , {\hat P}_\nu] = i g_{\mu \nu} \hbar F \left(\frac{{\hat x_{\mu}}}{x_m}, \frac{{\hat p_{\mu}}}{p_M} \right) ~.
\label{eq:modcomm-1}
\end{equation}
The indices $\mu , \nu$ take on the values $0, 1, 2, 3$,  and $g_{\mu \nu} = diag (1, -1, -1, -1)$ is the $3+1$ flat space--time metric. 

A more general version of the commutator from \eqref{eq:modcomm-1} would be a modified commutator of the form $[{\hat X}_\mu , {\hat P}_\nu] = i  \hbar F _{\mu \nu} \left(\frac{{\hat x_{\mu}}}{x_m}, \frac{{\hat p_{\mu}}}{p_M} \right)$, where we modify each component of the metric separately rather than having an overall scalar modification of the commutator. In the limit limit of large distance/small momentum, one would need to have the limit $F_{\mu \nu} \to g_{\mu \nu}$. We made the choice of a scalar modification to the commutator given in \eqref{eq:modcomm-1} primarily for simplicity. It would be worth considering a more general modification, such as where one modifies each component of the metric separately. 

Next, we define modified space--time operators, ${\hat X}_\mu$, and energy--momentum operators, ${\hat P}_\mu$, which will yield \eqref{eq:modcomm-1}. The modified position operator is written as
\begin{equation}
    \label{x3d}
{\hat X}_\mu = i \hbar G\left(\frac{{\hat x}_\nu}{x_m}, \frac{{\hat p}_\nu}{p_M} \right) \partial _{p_\mu}  ~~~~\to ~~~~~~ {\hat X}_\mu =  i \hbar G\left(\frac{p}{p_M} \right) \partial _{p_\mu} ~.
\end{equation}
In the last step, we assumed that the ansatz $G$ depends only on the magnitude of the canonical momentum, {\it, i.e.,}, $p = |p_\mu|$. Next, we take the modified momentum operator to be
\begin{equation}
    \label{p3d}
{\hat P}_\mu = H\left(\frac{{\hat x}_\nu}{x_m}, \frac{{\hat p}_\nu}{p_M} \right) {\hat p}_\mu ~~~~~ \to ~~~~~{\hat P}_\mu =  H\left(\frac{p}{p_M} \right) ~{\hat p}_\mu~,
\end{equation}
where again have taken the ansatz function $H$ to depend only on the magnitude of the canonical momentum. 
Now, $G$ and $H$ must to be chosen to yield  \eqref{eq:modcomm-1}. 

For both the modified position \eqref{x3d} and modified momentum \eqref{p3d} we assumed that the modification ansatz functions only depended on the magnitude of the momentum, (i.e., $p = |p_\mu|$) rather than on the momentum, (i.e., $p_\mu$). This was done to help ensure rotational symmetry. With a direct dependence on $p_\mu$, we were unable to easily preserve rotational~symmetry. 

Following reference \cite{BJLS}, we want to select the ansatz functions $G$ and $H$ in \eqref{x3d}~and~\eqref{p3d} so that $F=1$ 
in \eqref{eq:modcomm-1},  i.e., so that ${\hat X}_\mu$ and ${\hat P}_\mu$ satisfy the standard commutator relationship
\begin{equation}
[{\hat X}_\mu, {\hat P}_\nu]=i \hbar g_{\mu \nu}.   
\label{eq:cancomm-modvar}
\end{equation} 
The ansatz function $H$ is chosen to be the $3+1$-dimensional generalization of the 1-dimensional ansatz function from \eqref{ansatz-H}, namely \cite{plb-2023},
\begin{equation}
    \label{ansatz-h}
 H\left( \frac{p}{p_M} \right) = \frac{2 p_M}{\pi p} \arctan \left( \frac{\pi p }{2 p_M} \right)   ~~~~{\rm and}~~~~
  H\left( \frac{p }{p_M} \right) = \frac{p_M}{p} \tanh \left( \frac{p}{p_M} \right) ~.
\end{equation}
Although the ansatz function in \eqref{ansatz-h} appears identical to the one in \eqref{ansatz-H}, it should be noted that the function $H$ in \eqref{ansatz-h} depends on $p = |p_\mu| = \sqrt{p_0^2 - p_i ^2}$, not on the one-dimensional momentum as in \eqref{ansatz-H}. This distinction will come into play shortly when we examine the commutator of the modified position with the modified momentum. The $G$ ansatz functions which generalize the 1-dimensional version from \eqref{ansatz-G} are written as 
\begin{equation}
    \label{ansatz-g}
   G  \left( \frac{p}{p_M} \right) = \left[ 1+ \left( \frac{\pi p}{2 p_M}\right)^2 \right] ~~~~{\rm and}~~~~
   G  \left( \frac{p}{p_M} \right) = \cosh ^2 \left( \frac{p}{p_M}\right)  ~.
\end{equation}
Again, while this appears identical to the ansatz function $G$ from \eqref{ansatz-g}, the ansatz function $G$ in \eqref{ansatz-g} depends on the magnitude of the four-momentum rather than the magnitude of the one-dimensional momentum. 

Using \eqref{ansatz-h} and \eqref{ansatz-g}, the modified position and momentum operators for the $\tanh$ modification are 
\begin{equation}
    \label{xp3dc}
    {\hat X}_\mu = i \hbar \cosh ^2 \left( \frac{p}{p_M}\right) \partial_{p_\mu} ~~~~{\rm and}~~~~
     {\hat P}_\mu=  p_i~H \left( \frac{p}{p_M} \right) = p_\mu \frac{p_M}{p} \tanh \left( \frac{p}{p_M} \right) ~,
\end{equation}
and for the arctan modification
\begin{equation}
    \label{xp3dd}
    {\hat X}_\mu = i \hbar \left[ 1+ \left( \frac{\pi p}{2 p_M}\right)^2 \right] \partial_{p_\mu} 
    ~~~~{\rm and}~~~~
    {\hat P}_\mu = p_\mu~H \left( \frac{p}{p_M} \right) = p_\mu \frac{2 p_M}{\pi p} \arctan \left( \frac{\pi p}{2 p_M} \right) ~.
\end{equation} 

In order for the operators in \eqref{xp3dc} and \eqref{xp3dd} to satisfy \eqref{eq:cancomm-modvar}, the standard commutator, we require
\begin{equation}
\label{cond3d}
    G  \left( \frac{p}{p_M} \right) \partial_{p_\mu} \left[ p_\nu ~H \left(\frac{p}{p_M} \right)\right]  \stackrel{?}{=} g _{\mu \nu} ~.
\end{equation}
The question mark indicates that it is an open question whether the left-hand side can be equal to the right-hand side. 

For the modified momentum from \eqref{xp3dc}, condition \eqref{cond3d} becomes \cite{plb-2023}

\begin{eqnarray}
    \label{tanh-cond}
    G  \left( \frac{ p}{p_M} \right) \partial_{p_\mu} \left[ p_\nu ~H \left(\frac{p}{p_M} \right)\right] &=& \left[ \frac{p_M}{p} \sinh \left( \frac{p}{p_M} \right) \cosh \left( \frac{p}{p_M} \right)\left( g_{\mu \nu } - \frac{p_\mu p_\nu}{p^2} \right) + \frac{p_\mu p_\nu}{p^2} \right] \nonumber \\ &=& g_{\mu \nu} + \mathcal{O}\left(\frac{p^2}{p_M^2} \right) ~,
\end{eqnarray}

For the modified momentum from \eqref{xp3dd}, condition \eqref{cond3d} becomes \cite{plb-2023}

\begin{eqnarray}
    \label{arctan-cond}
    G  \left( \frac{p}{p_M} \right) \partial_{p_\mu} \left[ p_\nu ~H \left(\frac{p}{p_M} \right)\right] &=& \left[ \left( 1+ \left( \frac{\pi p}{2 p_M}\right)^2 \right) \frac{2 p_M}{\pi p} \arctan \left( \frac{\pi p}{2 p_M} \right)\left( g_{\mu \nu } - \frac{p_\mu p_\nu}{p^2} \right) + \frac{p_\mu p_\nu}{p^2} \right] \nonumber \\
    &=& g_{\mu \nu} + \mathcal{O}\left(\frac{p^2}{p_M^2} \right) ~.
\end{eqnarray}
In the second lines of \eqref{tanh-cond} and \eqref{arctan-cond}, the functions in square brackets on the right-hand side were expanded to the first order in $|{\vec p}|/p_M$. Equations \eqref{tanh-cond} and \eqref{arctan-cond} show that to the first order, one recovers the standard three-dimensional commutators. However, there is a difference at second order. In $3+1$ dimensions, it is no longer possible to modify the operators and retain the standard commutator. This is due to the fact that the ansatz functions now depend on $|p_\mu|$, which involves all the three spatial and one time component, whereas in the previous section the ansatz functions only depended on one momentum component. 

Using Equations \eqref{xp3dc} and \eqref{xp3dd} and the results of \eqref{tanh-cond}, we find that the modified commutator for the $\tanh$ modification, to the second order, is \cite{plb-2023}  
\begin{equation}
    \label{tanh-comm}
    [{\hat X}_\mu , {\hat P}_\nu] \approx i \hbar     
        \left[  g_{\mu \nu} + \frac{p^2}{2 p_M ^2} \left( g_{\mu \nu} - \frac{p_\mu p_\nu}{p^2} \right) \right]~.
\end{equation}

For the $\arctan$ modification, there is a subtle point. From \eqref{arctan-cond}, we want to expand a term of form $(1+x^2) \frac{\arctan (x)}{x}$, with $x \equiv \frac{\pi |{\vec p}|}{2 p_M}$. Here, it is necessary to expand $\arctan$ to ${\cal O} (x^3)$ due to $x$ in the denominator. For this order, we have $\arctan (x) \approx x - \frac{x^3}{3} $. Thus, for ${\cal O} (x^2)$, Equation \eqref{arctan-cond}  becomes  \cite{plb-2023}     
\begin{equation}
    \label{arctan-comm}
     [{\hat X}_\mu , {\hat P}_\nu] \approx i \hbar     \left[ g_{\mu \nu}+ \frac{2}{3} \left( \frac{\pi p}{2 p_M}\right)^2  \left( g_{\mu \nu} - \frac{p_\mu p_\nu}{p^2} \right)  \right] ~.
\end{equation}

The approximate commutators in \eqref{tanh-comm} and \eqref{arctan-comm} are similar to the modified commutators in \cite{KMM}---in both cases, the modified commutators are quadratic in $p$. However, unlike the one-dimensional case, the standard commutator \eqref{eq:cancomm-modvar} can no longer be recovered with the modified operators from \eqref{xp3dc} and \eqref{xp3dd}.

The results in \eqref{tanh-comm} and \eqref{arctan-comm} are good up to the second order in the ratio of momentum to cut-off momentum, $\frac{p^2}{p_M ^2}$. As the momentum approaches the cut-off momentum, $p \to p_M$, one would need to keep higher-order terms, which, in principle, could alter the results. It is not clear at this point whether these higher-order corrections would greatly alter the results presented here. We leave this question for possible future work.

\section{Modified Lorentz Group for Modified Operators}

In this section, we investigate what happens to the Lorentz symmetry with the modification of the operators given in \eqref{xp3dc} and \eqref{xp3dd}. In general, one expects that introducing a minimal length and/or modifying the space--time and energy--momentum operators will alter/break the original Lorentz symmetry. However, we find that it is possible to retain Lorentz symmetry despite the modification of the operators. 

First, we review the standard Lorentz symmetry for standard space--time and energy--momentum operators. One can construct the generators for the Lorentz group from the standard space--time and energy--momentum operators as
\begin{equation}
\label{LG-gen}
{\hat l}_{\alpha \beta} = {\hat x}_\alpha {\hat p}_\beta - {\hat x}_\beta {\hat p}_\alpha ~.
\end{equation}
For $\alpha =0$ and $\beta =i$, we have boost generators ${\hat k} _i = {\hat l}_{0i}$. For $\alpha =i$ and $\beta =j$, one has rotation generators, namely, angular momentum ${\hat l}_i = \frac{1}{2} \epsilon _{ijk} {\hat l_{jk}} = \epsilon_{ijk} {\hat x}_j {\hat p}_k$. These generators satisfy the standard Lorentz algebra, namely,
 \begin{equation}
    \label{lorentz-a}
     [{\hat l}_i, {\hat l}_j]=  i \hbar \epsilon _{ijk} {\hat l}_k ~~~ {\rm and} ~~~
     [{\hat k}_i, {\hat k}_j]= - i \hbar \epsilon _{ijk} {\hat k}_k ~~~ {\rm and} ~~~  [{\hat l}_i, {\hat k}_j]=  i \hbar \epsilon _{ijk} {\hat k}_k~.
 \end{equation}

Now, with the modified space--time and energy--momentum operators from \eqref{xp3dc} and \eqref{xp3dd}, one can construct modified generators of the Lorentz group as
\begin{equation}
\label{LG-gen-mod}
{\hat L}_{\alpha \beta} = {\hat X}_\alpha {\hat P}_\beta - {\hat X}_\beta {\hat P}_\alpha ~.
\end{equation}
These give new rotation generators of the form ${\hat L}_{ij} = {\hat X}_i {\hat P}_j - {\hat X}_j {\hat P}_i$, which can be written in vector form as ${\hat L}_i = \frac{1}{2} \epsilon _{ijk} {\hat L}_{jk} = \epsilon_{ijk} {\hat X}_j {\hat P}_k$. The new boost generators in vector form are ${\hat K}_i = {\hat L}_{0i} = {\hat X}_0 {\hat P}_i - {\hat X}_i {\hat P}_0$. 

It is easy to verify that these modified angular momentum operators and modified boost operators {\it do not} satisfy the standard algebra in \eqref{LG-gen} for the arctan modified operators or the tanh modified operators. 

However, following reference \cite{KMM}, it is possible to make simple modifications to the operators defined in \eqref{LG-gen-mod} that satisfy the Lorentz algebra of \eqref{LG-gen}. For the tanh modified operators, one can define a modified angular momentum operator as
 \begin{equation}
    \label{ang-mom-1}
     {\hat {\cal L}}_i \equiv \frac{p}{p_M \cosh ^2 ( p /p_M) \tanh (p/p_M)} {\hat L}_i 
  = \epsilon_{ijk} {\hat x}_j {\hat p}_k = {\hat l}_i~,
 \end{equation}
 and for the $\arctan$ modified operators from \eqref{xp3dd}, one has
 \begin{equation}
     \label{ang-mom-2}
     {\hat {\cal L}}_i = \frac{\pi p}{2 p_M [1+(\pi  p /2 p_M)^2 ] \arctan (\pi p/2 p_M)} {\hat L}_i  =  \epsilon_{ijk} {\hat x}_j {\hat p}_k = {\hat l}_i~.
 \end{equation}

Next, we turn to the modified boost generators which are the time--space components of ${\hat L}_{\alpha \beta}$, namely, ${\hat L}_{0i} = {\hat X}_0 {\hat P}_i - {\hat X}_i {\hat P}_0$. We define modified ``time'' and ``energy'' operators ${\hat X}_0 = {\hat T}$ and ${\hat P}_0 = {\hat {\cal E}}$ (note that we have set $c=1$ here and throughout the paper). To maintain the structure, the time operator needs to be modified via the same factors as the position operator, and the energy operator needs to be modified via the same factors as the momentum operator. For the $\tanh$ modification of \eqref{xp3dc}, this results in modified time and energy operators of the form
\begin{equation}
  \label{te3dc}
    {\hat X}_0 = {\hat T} = i \hbar  \cosh ^2 \left( \frac{E}{E_M}\right) \partial_{E} ~~~~{\rm and}~~~~
      {\hat P}_0 = {\hat {\cal E}} =  E_M \tanh \left( \frac{E}{E_M} \right) ~,
 \end{equation}
 where $E_M = p_M$, and $E=|p_\mu|=p$ is the usual energy operator. For the
 $\arctan$ modification this results in the  time and energy operators
 \begin{equation}
     \label{te3dd}
    {\hat X}_0 = {\hat T} = i \hbar \left[ 1+ \left( \frac{\pi E}{2 E_M}\right)^2 \right] \partial_E    ~~~~{\rm and}~~~~
   {\hat P}_0 = {\hat {\cal E}} =  \frac{2 E_M}{ \pi } \arctan \left( \frac{\pi E}{2 E_M} \right) ~.
 \end{equation}
Now, using the time and energy operators from \eqref{te3dc} and \eqref{te3dd}, we define new boost operators analogously to the new rotation operators of \eqref{ang-mom-1} and \eqref{ang-mom-2}. For the $\tanh$ modification, we have
  \begin{eqnarray}
    \label{boost-1}
     {\hat {\cal K}}_i &\equiv& \frac{E}{E_M \cosh ^2 ( E /E_M) \tanh (E/E_M)} {\hat K}_i \nonumber \\
     &=& \frac{p}{p_M \cosh ^2 ( p /p_M) \tanh (p/p_M)} {\hat K}_i = {\hat k}_i~,
 \end{eqnarray}
 and for the $\arctan$ modified operators from \eqref{xp3dd}, one has
 \begin{eqnarray}
     \label{boost-2}
     {\hat {\cal K}}_i &\equiv& \frac{\pi E}{2 E_M [1+(\pi  E /2 E_M)^2 ] \arctan (\pi E/2 E_M)} {\hat K}_i \nonumber \\
     &=& \frac{\pi p}{2 p_M [1+(\pi  p /2 p_M)^2 ] \arctan (\pi p/2 p_M)} {\hat K}_i  = {\hat k}_i~.
 \end{eqnarray}
In the second lines of \eqref{boost-1} and \eqref{boost-2}, we have used the fact that $E_M = p_M$ and $E=|p_\mu| = p$. Looking at the newly defined angular momentum and boost operators in \eqref{ang-mom-1}, \eqref{ang-mom-2} and \eqref{boost-1}, and \eqref{boost-2}, we see that they are equivalent to the standard operators. Thus, these new operators will satisfy the same standard Lorentz algebra from \eqref{LG-gen}

The modified angular momentum and boost operators defined in \eqref{ang-mom-1} and \eqref{boost-1} or \eqref{ang-mom-2} and \eqref{boost-2} trivially satisfy the same Lorentz algebra \eqref{lorentz-a} as the standard operators
 \begin{equation}
    \label{lorentz-aaa}
     [{\hat {\cal L}}_i, {\hat {\cal L}}_j]=  i \hbar \epsilon _{ijk} {\hat {\cal L}}_k ~~~ {\rm and} ~~~
     [{\hat {\cal K}}_i, {\hat {\cal K}}_j]= - i \hbar \epsilon _{ijk} {\hat {\cal K}}_k ~~~ {\rm and} ~~~  [{\hat {\cal L}}_i, {\hat {\cal K}}_j]=  i \hbar \epsilon _{ijk} {\hat {\cal  K}}_k~.
 \end{equation}
Thus, although the modifications of the time--space and energy--momentum operators lead to a minimal length---and more specifically a minimum length that is different for directions along the momentum of the system versus orthogonal to the direction of the momentum---they do so without violating the Lorentz algebra, at least in regard to rotations and boosts. This is closely related to the GUP model in \cite{BJLS, grf-2022}, which defined modified operators that resulted in a minimal length but still preserved the standard special relativistic relationship between energy and momentum. Retaining the standard special relativistic energy--momentum dispersion relationship for the modified energy and momentum allowed one to have a theory with a minimal length but which did not run afoul of the observational results from short gamma-ray bursts \cite{grb-1,grb-2}, which showed no sign of photon dispersion up to scales shorter than the Planck distance. Generic GUP models usually have photon dispersion, which therefore provides a possible way to test such GUP models. The GUP models in this section do not exhibit photon dispersion and thus are not testable via observations of short gamma-ray bursts.   

Although the Lorentz algebra is preserved for the modified rotation and boost generators, as shown in \eqref{lorentz-a}, one can ask whether there are physical consequences that could distinguish this GUP model with its Lorentz algebra, but modified position and momentum operators from the standard Lorentz algebra with standard position and momentum operators. As mentioned above, the present GUP model does not exhibit photon dispersion, as is the case for the standard Lorentz algebra. However, in references \cite{BJLS,grf-2022} it was suggested that these modified GUPs could offer an explanation for the appearance of ultra-high-energy cosmic rays beyond the GZK \cite{gzk-1,gzk-2} cut-off.  

Additionally, the present GUP models do have some deviation from the usual Lorentz algebra. In the standard Lorentz algebra, position operators commute with other position operators, (i.e., $[{\hat x}_i, {\hat x}_j]=0$), and momentum operators commute among themselves, (i.e., $[{\hat p}_i, {\hat p}_j]=0$). In the next section, we show that while the modified momentum operators still commute, the modified position operators do not. This leads to a connection between GUPs and non-commutative geometry, another approach to quantum gravity. 

\section{3D GUP and Non-Commutative Geometry}

In moving from one-dimensional to three-dimensional GUP models, one can see there is a congruence to the field of non-commutative geometries, which is a different (but in the end hopefully equivalent) approach to a minimal distance scale in phenomenological quantum gravity. A review of non-commutative geometry can be found in \cite{piero}. Non-commutative geometry models are most often used to smooth out the singularities in black holes \cite{euro}, whereas GUP models tend to focus on the high-energy/high-momentum regime of particles. Non-commutative geometries approach quantum gravity from the gravity side, while the GUP approaches quantum gravity from the quantum field theory side. 

The connection between the 3D GUP and non-commutative geometry models was pointed out in the early work \cite{KMM}. However, to the best of our knowledge, no one has made a detailed analysis of the connections between minimal lengths achieved through non-commutative geometry and those achieved through 3D GUP models and how consistent the models are with one another. In this section, we take the first step in this direction. 

 One can see that the three-dimensional modified position operators of Equations \eqref{xp3dc} and \eqref{xp3dd} imply a non-commutativity between the coordinates. From \eqref{xp3dc}, one finds
 \begin{eqnarray}
 \label{xx-tanh}
     [{\hat X}_i, {\hat X}_j] &=& \frac{2 i \hbar \cosh ^3(|{\vec p}|/p_M) \sinh (|{\vec p}|/p_M)}{p_M |{\vec p}|} ({\hat p}_i {\hat x}_j - {\hat p}_j {\hat x}_i) \nonumber \\
  &=& \frac{2 i \hbar \cosh (|{\vec p}|/p_M) \sinh (|{\vec p}|/p_M)}{p_M |{\vec p}|} ({\hat p}_i {\hat X}_j - {\hat p}_j {\hat X}_i)  \\
    &=& \frac{2 i \hbar \cosh ^2(|{\vec p}|/p_M)}{p_M^2} ({\hat P}_i {\hat X}_j - {\hat P}_j {\hat X}_i) \nonumber \\
     &=& \frac{2 i \hbar}{p_M ^2 - |{\vec P}|^2} ({\hat P}_i {\hat X}_j - {\hat P}_j {\hat X}_i)~. \nonumber
 \end{eqnarray}
 The first line in \eqref{xx-tanh} gives the right hand side of the commutator in terms of the standard position and momentum operators. In the last line, we have given the right hand side of the commutator in terms of the modified position and momentum operators by changing $|{\vec p}|$ to $|{\vec P}|$ and by solving the connection between the two momenta given in \eqref{xp3dc}, namely, $\frac{|{\vec p}|}{p_M} = \tanh ^{-1} (|\vec{P}| / p_M)$. Using this relationship in the third line of \eqref{xx-tanh} then gives the last line in \eqref{xx-tanh}, which is entirely in terms of the modified operators.

For the $\arctan$ modified position of \eqref{xp3dd}, we obtain
 \begin{eqnarray}
 \label{xx-arctan}
     [{\hat X}_i, {\hat X}_j] &=& 2 i \hbar \left[ 1+ \left(\frac{\pi |{\vec p}|}{2 p_M}\right)^2 \right] \left( \frac{\pi}{2 p_M} \right)^2  ({\hat p}_i {\hat x}_j - {\hat p}_j {\hat x}_i) \nonumber \\
    &=& 2 i \hbar \left( \frac{\pi}{2 p_M} \right)^2  ({\hat p}_i {\hat X}_j - {\hat p}_j {\hat X}_i) \\
    &=& \frac{2 i \hbar}{\arctan (\pi |{\vec p}|/2 p_M)} \left( \frac{\pi}{2 p_M} \right)^2 \left( \frac{\pi |{\vec p}|}{2 p_M}\right) ({\hat P}_i {\hat X}_j - {\hat P}_j {\hat X}_i) \nonumber \\
     &=& 2 i \hbar \left( \frac{\pi}{2 p_M |{\vec P}|} \right) \tan \left( \frac{\pi |{\vec P}|}{2 p_M} \right) ({\hat P}_i {\hat X}_j - {\hat P}_j {\hat X}_i) \nonumber ~.
 \end{eqnarray}
Again, in the last step, we have replaced standard operators with the modified operator and converted $|{\vec p}|$ to $|{\vec P}|$ by solving the connection between the two momenta given in \eqref{xp3dc}, namely, $\frac{\pi |{\vec p}|}{2 p_M} = \tan (\pi |\vec{P}| / 2 p_M)$. When the modified momentum approaches its asymptotic value, $p_M$, the commutators in both \eqref{xx-tanh} and \eqref{xx-arctan} diverge.

The right-hand side of the modified position commutators in both \eqref{xx-tanh} and \eqref{xx-arctan} have the angular momentum operator on the right hand side---either the standard angular momentum operators in terms of the standard momentum and standard position (${\hat l}_{ij} = {\hat p}_i {\hat x}_j - {\hat p}_j {\hat x}_i$) or modified angular momentum operators in terms of the modified momentum and modified position (${\hat L}_{ij} = {\hat P}_i {\hat X}_j - {\hat P}_j {\hat X}_i$)---times some multiplicative factors that are momentum-dependent. All these features are equivalent to  the form of modified position commutators and non-commutative geometry that emerged from the GUP model in \cite{KMM}. However, \eqref{xx-tanh} and \eqref{xx-arctan} are both different from the standard non-commutative geometry discussed in \cite{piero}, which generally started with a position commutator of the form
\begin{equation}
     \label{nc-geo}
     [X_i, X_j] = i \Theta _{ij} ~,
 \end{equation} 
where $\Theta _{ij}$ is a constant, anti-symmetric tensor. Generally, the position commutator in standard non-commutative geometry is written with four-vector indices, $[X_\mu, X_\nu] = i \Theta _{\mu \nu}$, but here we use only three-vector indices to make the comparison with \eqref{xx-tanh} and
 \eqref{xx-arctan} more transparent. In contrast to the standard right-hand side of a position commutator like \eqref{nc-geo}, the right hand side of our modified position commutator, as in \eqref{xx-tanh} or \eqref{xx-arctan}, depends on momentum---either standard momentum or modified momentum.  

The commutators in \eqref{xx-tanh} and \eqref{xx-arctan} lead to a non-trivial uncertainty relationship between the modified position operators. For \eqref{xx-tanh}, the uncertainty relationship is
 \begin{eqnarray}
 \label{xx-up}
      \Delta  X_i \Delta X_j &\ge& \frac{\hbar}{p_M} \left\langle \frac{\cosh ^3(|{\vec p}|/p_M) \sinh (|{\vec p}|/p_M)}{|{\vec p}|} ({\hat p}_i {\hat x}_j - {\hat p}_j {\hat x}_i) \right \rangle \nonumber \\
      &\ge& \frac{\hbar}{p_M ^2} \left \langle ({\hat p}_i {\hat x}_j - {\hat p}_j {\hat x}_i) \right \rangle = \frac{\hbar}{p_M ^2} \epsilon _{ijk} \langle {\hat l}_k \rangle ~,
 \end{eqnarray}
where in the second line we have approximated the hyperbolic functions to the first order in $|{\vec p}|/p_M$ and written the expectation in terms of the ordinary angular momenta $\epsilon _{ijk} {\hat l}_k= {\hat p}_i {\hat x}_j - {\hat p}_j {\hat x}_i$. Since $\langle {\hat l}_k \rangle = n \frac{\hbar}{2}$---an integer multiple of $\hbar /2$---we find that the ``area'' uncertainty implied by \eqref{xx-up} is $|\Delta  X_i, \Delta X_j | \sim \frac{n \hbar ^2}{2 p_M ^2}$. This is consistent, up to factors of order unity, with what we found in Section 4 for the minimum uncertainty of the modified position. From Section 4, we found that for the same order in $|{\vec p}|/p_M$, $|\Delta X_i| \sim \frac {\hbar}{p_M}$, which then gave $|\Delta X_i \Delta X_j| \sim \frac {\hbar ^2}{p_M ^2}$ in general agreement with \eqref{xx-up}. 

The same results follow from \eqref{xx-arctan}, which yields an uncertainty relationship of
 \begin{eqnarray}
 \label{xx-up-2}
      \Delta  X_i \Delta X_j &\ge& \frac{\hbar \pi ^2}{4 p_M ^2} \left\langle  \left[ 1+ \left(\frac{\pi |{\vec p}|}{2 p_M}\right)^2 \right] ({\hat p}_i {\hat x}_j - {\hat p}_j {\hat x}_i) \right \rangle \nonumber \\
      &\ge& \frac{\hbar \pi ^2}{4 p_M ^2} \left \langle ({\hat p}_i {\hat x}_j - {\hat p}_j {\hat x}_i) \right \rangle = \frac{\hbar \pi ^2}{4 p_M ^2} \epsilon _{ijk} \langle {\hat l}_k \rangle ~,
\end{eqnarray}
where in the second line we have again approximated the expression to the first order in $|{\vec p}|/p_M$ and inserted the angular momentum operator.  Again, the area uncertainty implied by \eqref{xx-up-2}, (i.e., $|\Delta  X_i \Delta X_j| \sim \frac{n \hbar ^2 \pi ^2}{8 p_M ^2}$) is consistent with the implied ``area'' uncertainty from Section 4 obtained via GUP arguments. 

One final comment is that both \eqref{xx-up} and \eqref{xx-up-2} imply a relationship between the area uncertainty and the angular momentum. This connection is also found in spin foam models of quantum gravity \cite{baez,alejandro}, where one has that the following relationship between the area operator ${\hat A}_S$ for surface $S$ is given by
\begin{equation}
    \label{s-foam}
    {\hat A}_S | \Psi \rangle = 8 \pi l_{Pl} ^2 \gamma\sqrt{l(l+1)} | \Psi \rangle ~,
\end{equation}
where $l_{Pl}$ is the Planck length and $\gamma$ is the Immirzi parameter \cite{immirzi}. The spin foam relationship between the area operator and the angular momentum in \eqref{s-foam} is similar to the relationship in \eqref{xx-up} and \eqref{xx-up-2}. The area operator on the left hand side of \eqref{s-foam} is simply the magnitude of the area, while the ``areas'' in \eqref{xx-up} or \eqref{xx-up-2} retain the vector area nature through the indices on $\Delta X _i$. Similarly, the right-hand side of \eqref{s-foam} deals with the magnitude of the angular momentum operator, (i.e., $\sqrt{l(l+1)} \sim \sqrt{\frac{{\hat l}^2}{\hbar ^2}}$), while the right-hand side of \eqref{xx-up} or \eqref{xx-up-2} depends on the expectation of the angular momentum operator, (i.e., $\langle {\hat l}_k \rangle$). If one takes the the cut-off momentum equal to the Planck momentum, $p_M \to p_{Pl}$, recalls the relationship between between Planck momentum and Planck length ($p_{Pl} = \frac{\hbar}{l_{Pl}}$), and sets $\langle {\hat l}_k \rangle \sim l \hbar$ (where $l$ is an integer of order unity), then one finds that the right-hand sides of \eqref{xx-up} and \eqref{xx-up-2} are approximately equivalent to the right hand side of \eqref{s-foam}. Aside from the vector character \mbox{of \eqref{xx-up}} and \eqref{xx-up-2} versus the scalar character of \eqref{s-foam}, one can see the close connection between the non-commutative uncertainty relationships of positions with the area--angular momentum relationship of spin foam models. 
 
\section{Summary and Conclusions}

In this paper, we have extended previous work on 1D GUP models \cite{aiken-2019,BLS,BJLS} to 3D. Our aim in the work on the 1D models was to construct a GUP model with a minimal length, which, while modifying both position and momentum operators, left the commutator the same as in standard quantum mechanics. 
Further in \cite{BJLS}, we wanted to give a GUP model with a minimal length which did not spoil the energy--momentum dispersion relationship, $E^2-p^2 = m^2$, thus avoiding the non-observation of GUP effects like photon dispersion \cite{AC-nature,grb-1,grb-2}.
Section \ref{sec2}, summarized the 1D GUP models. Section \ref{sec3} extended these 1D models to 3D while retaining rotational invariance. This was accomplished by requiring the modifications to be functions of the total momentum $|\vec{p}|$.  
Consequently, we found that the 3D commutator had to take a form different from the standard quantum mechanical commutator in at least one direction. 
This is seen explicitly in Equations \eqref{tanh-comm} and \eqref{arctan-comm}, where the second-order corrections to the position-momentum commutators were of the form $\propto \delta_{ij} - \frac{p_j p_i}{|{\vec p}|^2}$. In 1D GUP models, the term $\delta_{ij} - \frac{p_j p_i}{|{\vec p}|^2}$ is zero. 
We investigated the uncertainty relationships arising from these modified 3D operators and modified commutators. 
We found that the minimum uncertainty in position was approximately $\Delta X_i ^{min} \sim \frac{\hbar}{p_M}$ \cite{plb-2023}---the same result as in 1D. 

Next, we examined the minimum distance in two cases.
First, we used spherically symmetric wave functions to analyze how small a volume the wave function could be compressed. We found it to be the minimal distance cubed up to a factor of order unity.  
Second, we examined a system with a large, nonzero momentum in one direction. We found that the minimum distance in the direction of the average momentum was different from the minimum distance in the two orthogonal directions. For the $\tanh$ and $\arctan$ GUPs, the positional uncertainty in the orthogonal directions was greater than in the direction of nonzero momentum. A similar result occurred in the GUP model in reference \cite{KMM}, i.e., $\Delta X_{2,3} ^{min} > \Delta X_1 ^{min}$.

The difference between the minimum position uncertainties in the second case above points to the possibility of a violation of Lorentz symmetry. The existence of minimal distances already naively implies a violation of Lorentz symmetry. 
However, in \cite{BJLS}, a 1D GUP model was formulated, which gave a minimal length while preserving the special relativistic energy-momentum relationship, $E^2 - |{\vec p}|^2 = m^2$, thus preserving Lorentz symmetry \cite{plb-2023}.  A related issue is the connection of the present work to breaking of the isotropy of space, as observed for other GUP models \cite{husin}. We plan to study the violation of Lorentz symmetry of these modified position and momentum operators in future work. 

The generators of the canonical Lorentz algebra
are angular momentum, (i.e.,  ${\hat l}_{ij} = {\hat x}_i {\hat p}_j - {\hat x}_j {\hat p}_i$) and boost operators, (i.e., ${\hat l}_{0i} = -{\hat x}_0 {\hat p}_i + {\hat x}_i {\hat p}_0$). 
The central question is then ``How do the modified position and momentum operators affect the observables associated with modified angular momentum and boost operators?". 
Modified boost operators are constructed by defining modified time and energy operators based on the $\tanh$ or $\arctan$ operators of equations \eqref{xp3dc} and \eqref{xp3dd}.

Another fascinating consequence of the modifications is the new commutation relations of position operators among themselves. 
The standard result is $[{\hat x}_i, {\hat x}_j] =  0$; the modified position operators from \eqref{xp3dc} and \eqref{xp3dd} satisfy $[{\hat X}_i, {\hat X}_j] \propto \epsilon_{ijk}  {\hat l}_k$ so that the commutators are proportional to the angular momentum operators.
The non-commutativity of the modified position operators provides a link between GUP models and non-commutative geometry models \cite{piero,euro}.
Consequently, the related uncertainty relationships are $\Delta X_i \Delta X_j \propto  \epsilon_{ijk}  \langle  {\hat l}_k\rangle $ and the uncertainty of the cross-sectional area is connected to the expectation of the angular momentum operator. 
This provides a potentially dynamic non-commutative geometry that warrants further study.  
Moreover, this provides another connection between the 3D GUP models in this work and spin foam models \cite{baez,alejandro}.
 
We leave these questions---the modification of the Lorentz algebra and the connection of 3D GUP models to non-commutative geometry and spin foam models---for future work. 

Finally, in this work, we have not addressed the ``soccer ball'' problem of GUP models, i.e., how this applies to multi-particle states and/or fields \cite{hossenfelder}. We leave this difficult and currently unresolved problem for future work.

\end{document}